\documentclass[twocolumn,secnumarabic,amsmath,amssymb,nobibnotes,aps,prd]{revtex4-2}
\usepackage{graphicx}
\begin{document}
	\title{Direct solution of Minkowski-space Bethe-Salpeter equation in the massive Wick-Cutkosky model}
	\author{Shaoyang Jia}
	\email[]{syjia@anl.gov}
	\affiliation{Physics Division, Argonne National Laboratory, 9700 South Cass Avenue, Lemont, Illinois 60439, USA}
	\date{February 20, 2024}
	\begin{abstract}
		In order to solve the Bethe-Salpeter equation (BSE) in the Minkowski space, we first introduce the Nakanishi integral representations of the Bethe-Salpeter amplitude (BSA) and the Bethe-Salpeter wave function (BSWF). We then derive the explicit integral equations for the corresponding spectral functions from the BSE for states of $2$ scalar particles bound by a scalar-particle exchange interaction, where the propagators of constituents are allowed to be fully dressed. These integral equations are subsequently solved numerically in the variation of the Wick-Cutkosky model with massive exchange particles, where an algorithm of adaptive mash grid is proposed. The equations and algorithm we develop here serve as the foundation of Minkowski-space formulation of BSE for bound states of fermions.
	\end{abstract}
	\maketitle
	\section{Introduction\label{sc:introduction}}
	In terms of Green's functions of a quantum field theory, the structure of a $2$-body bound bound state is described by its Bethe-Salpeter amplitude (BSA), which is determined from the Bethe-Salpeter equation (BSE). One main advantage of this method is its support of direct formulation in the Minkowski space, especially useful in the evaluation of inelastic properties of the bound state. Inputs to the BSE are propagators of the constituents and vertices of their interactions, both of which can also be formulated in the Minkowski space~\cite{Kusaka:1995za,Kusaka:1997xd}. The introduction of integral representations facilitates the solution of propagators from their Schwinger-Dyson equation~\cite{Jia:2017niz} and the BSAs from their BSEs. These representations are introduced to describe the Green's functions in terms of their discontinuities at the boundaries of their holomorphic region. For scalar particles, they are the K\"all\'en-Lehmann representation for the propagators and the Nakanishi integral representation for the BSAs~\cite{nakanishi1971graph,Kusaka:1995za,Kusaka:1997xd}. While one generalization to fermion propagators and the fermion-photon vertex in quantum electrodynamics (QED) is the gauge technique~\cite{Delbourgo:1977jc,Salam:1963sa,Salam:1964zk,Strathdee:1964zz}. 
	
	When only the regions of spacelike momenta are of interest, one well-established method of solving for the BSEs utilizes the Wick rotation~\cite{Maris:1999nt,Fischer:2014xha,Eichmann:2021vnj}. Because singularities are not expected to present in these regions, BSAs are solved directly from their BSEs using iterative solvers. While the Nakanishi integral representation fills in the gap when information about the BSA with timelike momenta are requested~\cite{nakanishi1971graph}. Although indirect methods of finding the Nakanishi spectral functions exist~\cite{Karmanov:2005nv,Carbonell:2005nw,Frederico:2013vga,dePaula:2016oct,Gutierrez:2016ixt,Karmanov:2021imh}, the BSE for scalar particles can be formulated as integral equations of Nakanishi spectral functions directly with numerical kernels of integration~\cite{Kusaka:1995za,Kusaka:1997xd}. Instead of numerical kernels, we convert the BSE for two scalar particles bound by a scalar exchange into integral equations of Nakanishi spectral functions of the BSA analytically.
	
	Within this article, we first introduce the Nakanishi integral representations for both the BSA and the Bethe-Salpeter wave function (BSWF). Because the K\"all\'en-Lehmann representation allows the propagators of scalar particles to be fully dressed in the Minkowski space, we then convert the momentum-space definition of the BSWF into an integral relation that merges the spectral functions of the propagators and of the BSA into that of the BSWF. In the meantime, the BSE provides another integral relation that evaluates the Nakanishi spectral function of the BSA in terms of the BSWF. These two integral identities form a closed system of equations required by an iterative solver. We then introduce an algorithm that solves the coupled integral equations with an adaptive mesh grid for spectral variables. We test these equations and the proposed algorithm in the massive variation of the Wick-Cutkosky model where the propagators of the constituents and the exchange particles are bare. 
	
	This article is organized as follows. Section~\ref{sc:introduction} is the introduction. Section~\ref{sc:nakanishi} shows the derivation of explicit integral equations for the Nakanishi spectral functions. Section~\ref{sc:numeric} contains the adaptive algorithm for the integral equations derived, the numerical settings, and the illustration of solution in the massive variation of the Wick-Cutkosky model. Section~\ref{sc:summary} gives the summary and concluding remarks. 
	\section{Nakanishi integral representation for BSE\label{sc:nakanishi}}
	\subsection{Representations of dressed propagators and Bethe-Salpeter amplitudes\label{ss:nakanishi_def}}
	Because we work in the Minkowski space, we adopt $g^{\mu\nu} = {\mathrm{diag}\{ 1, -1, -1, -1 \}}$ as the metric with ${3+1}$ spacetime dimensions. This convention results in ${q^2 \geq 0}$ for timelike momentum $q^\mu$. As an introduction to integral representations of Green's functions in the Minkowski space, let us start with the K\"all\'en-Lehmann spectral representation for the dressed propagator $D(p^2)$ of a scalar particle~\cite{Kusaka:1995za}. Explicitly we have the following integral representation:
	\begin{equation}
		D(p^2)=\int_{s_{\mathrm{th}}}^{+\infty}ds\,\dfrac{\rho(s)}{p^2-s+i\varepsilon}, \label{eq:sr_scalar_prop}
	\end{equation}
	where $\rho(s)$ is the spectral function of the propagator. Specifically when $\rho(s)=\delta(s-\mu^2)$, we reproduce the free-particle propagator for particles of mass $\mu$:
	\begin{equation}
		D_{0}(p^2) = 1/(p^2-\mu^2+i\varepsilon). \label{eq:def_bare_prop}
	\end{equation}
	The $i\varepsilon$ term in the denominators stands for the Feynman prescription, specifying the selection rules for poles when the temporal components of the momentum is integrated. Equation~\eqref{eq:sr_scalar_prop} consequently corresponds to propagators whose singularities in the complex momentum plane are given by poles and branch cuts in the timelike real axis starting from the threshold $p^2 \geq s_{\mathrm{th}}$. Such a representation has its generalization to fermion propagators in QED~\cite{Jia:2017niz}, which facilitates the solution of the gauge dependence for these propagators in the momentum space with arbitrary numbers of spacetime dimensions~\cite{Jia:2016udu,Jia:2016wyu}. 
	
	In terms of Green's functions, the structure of a scalar bound state of two scalar constituents is given by its BSA ${\psi(k^2,k\cdot P)}$~\cite{Maris:1999nt}. Kinematically $P^\mu$ is the momentum of the bound state. While those of the constituents are given by $k^\mu_\pm = {k^\mu \pm\eta_{\pm}P^\mu}$, with $\eta_{\pm}$ specifying the momentum partition. Momentum conservation then requires that $P^\mu = {k^\mu_+ - k^\mu_-}$, which indicates ${\eta_+ +\eta_-} = 1$. In this article, we work with ${\eta_+ = \eta_- = 1/2}$ by default. The Minkowski-space structure of the BSA is captured by the Nakanishi integral representation in the form of
	\begin{equation}
		\psi(k^2,k\cdot P)= \int_{-1}^{1}dz\int_{g_{\mathrm{th}}(z)}^{+\infty} d\gamma\,\dfrac{ \phi(\gamma,z) }{ ( k^2+zk\cdot P -\gamma+i\varepsilon )^n }, \label{eq:NIR_amplitude}
	\end{equation}
	where $\phi(\gamma,z)$ is the Nakanishi spectral function~\cite{nakanishi1971graph,Kusaka:1995za,Kusaka:1997xd}. We have allowed the radial threshold to depend on the angular variable $z$. The BSA given by Eq.~\eqref{eq:NIR_amplitude} is regular when both $k^2_+ < g_{\mathrm{th}}(z)$ and $k^2_- < g_{\mathrm{th}}(z)$. It encounters branch cuts when either $k^2_+\geq g_{\mathrm{th}}(z)$ or $k^2_-\geq g_{\mathrm{th}}(z)$. This can be illustrated by comparing Eq.~\eqref{eq:NIR_amplitude} with the integral representation of Appell $F_1$ function~\cite{10.1063/PT.3.3846}. Similar to the case of the propagator, the $i\varepsilon$ term in Eq.~\eqref{eq:NIR_amplitude} is the Feynman prescription. The power index $n$ of the denominator is tied to the degree of singularity in the BSA. For practical purposes, we find ${n=1}$ to be sufficient in numerical studies. 
	
	\begin{figure}
		\centering
		\includegraphics[width=\linewidth]{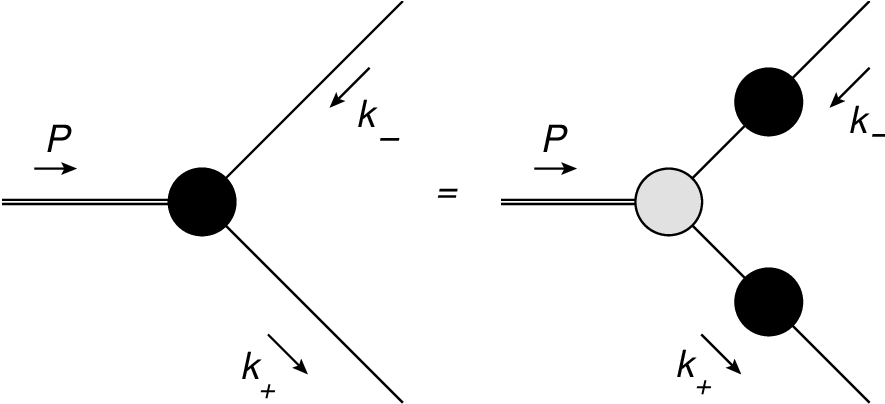}
		\caption{The Feynman diagrams representing Eq.~\eqref{eq:def_BSWF} as the definition of the BSWF in terms of the BSA and propagators. The black blobs are the BSWF and the dressed propagators. While the gray blob is the BSA. }\label{fig:bswf_def}
	\end{figure}
	The BSWF is a product of the BSA with propagators of its constituents. In the momentum space, we have
	\begin{equation}
		\chi(k^2,k\cdot P) = D_+(k^2_+)\,\psi( k^2,k\cdot P)\, D_-(k^2_-), \label{eq:def_BSWF}
	\end{equation}
	with ${\chi(k^2,k\cdot P)}$ being the BSWF~\cite{Maris:1999nt}. Functions ${D_\pm( k^2_\pm)}$ are the propagators of the constituents, whose spectral functions are ${\rho_\pm(s)}$. This relation is diagrammatically represented in Fig.~\ref{fig:bswf_def}. Similar to Eq.~\eqref{eq:NIR_amplitude}, the Nakanishi integral representation of the BSWF is given by
	\begin{align}
		& \chi(k^2,k\cdot P) \nonumber\\
		& = \int_{-1}^{+1}dz \int_{\Gamma_{\mathrm{th}}(z)}^{+\infty} d\gamma \dfrac{ \Phi(\gamma,z) }{ ( k^2 + zk\cdot P - \gamma +i\varepsilon )^{n+2} }, \label{eq:NIR_chi}
	\end{align}
	with $\Phi(\gamma,z)$ being the corresponding Nakanishi spectral function. Equation~\eqref{eq:NIR_chi} gives a Green's function with singularities similar to those discussed for the BSA in Eq.~\eqref{eq:NIR_amplitude}. Here the threshold ${\Gamma_{\mathrm{th}}(z)}$ is not necessarily the same as that in Eq.~\eqref{eq:NIR_amplitude}. The power index for the denominator in Eq.~\eqref{eq:NIR_chi} is chosen to be ${n+2}$ in order to match the power indices of momenta in Eqs.~\eqref{eq:sr_scalar_prop}~and~\eqref{eq:NIR_amplitude}. As reproducing the bare propagators through the K\"all\'en-Lehmann spectral representation requires $\delta$-functions, let us note that spectral functions of the propagator, of the BSA, and of the BSWF are not restricted to regular functions, but allowed to be distributions. 
	\subsection{Merging propagators with Bethe-Salpeter amplitude\label{ss:BSWF_merge}}
	The momentum-space definition of the BSWF in Eq.~\eqref{eq:def_BSWF} indicates an expression of the Nakanishi spectral function $\Phi(\gamma,z)$ in terms of spectral functions of the propagators and of the BSA. To derive this expression, let us start by applying the spectral representation of the propagators in Eq.~\eqref{eq:sr_scalar_prop} and the Nakanishi integral representation of the BSA in Eq.~\eqref{eq:NIR_amplitude}. Equation~\eqref{eq:def_BSWF} subsequently becomes
	\begin{align}
		& \chi(k^2,k\cdot P) = \int ds \int dt \int dz' \int d\gamma' \,\rho_+(s) \rho_-(t) \phi(\gamma', z') \nonumber\\
		& \quad \times \dfrac{1}{ ( k^2_+-s+i\varepsilon ) ( k^2_--t+i\varepsilon ) ( k^2+z'k\cdot P-\gamma'+i\varepsilon )^n } \nonumber\\
		& = n(n+1) \int ds \int dt \int dz' \int d\gamma' \, \rho_+(s) \rho_-(t) \,\phi(\gamma', z') \nonumber\\
		& \quad \times \int_{0}^{1}dx\int_{0}^{1-x}dy\, \dfrac{\left(1-x-y \right)^{n-1}}{ \mathbb{D}^{n+2}(x,y,z',s,t,\gamma',k^2,k\cdot P) },\label{eq:BSWF_Feynman_para}
	\end{align}
	where integration limits of spectral variables are implicit. Recall ${\rho_\pm(s)}$ are the spectral functions for the propagators of constituent particles. Here we have applied the Feynman parametrization to combine denominators into
	\begin{align*}
		& \mathbb{D}(x,y,z',s,t,\gamma',k^2,k\cdot P) = k^2 + \big[ (x-y)+(1-x-y) \nonumber\\
		& \times z' \big] k\cdot P - \big[ xs+yt+(1-x-y)\gamma'-(x+y)P^2/4 \big].
	\end{align*}
	Comparing the momentum dependence in Eq.~\eqref{eq:NIR_chi} with that in Eq.~\eqref{eq:BSWF_Feynman_para} gives
	\begin{align}
		& \Phi(\gamma,z) = n(n+1)\int ds\int dt \int dz'\int d\gamma' \,\rho_+(s)\,\rho_-(t)\, \nonumber\\
		& \quad \times \phi(\gamma',z')\int_{0}^{1}dx\int_{0}^{1-x}dy \, (1-x-y)^{n-1} \nonumber\\
		& \quad \times \delta\left( x-y+(1-x-y)z'-z \right) \nonumber\\
		& \quad \times \delta\left( xs + yt + (1-x-y)\gamma'- (x+y)P^2/4 - \gamma \right)\label{eq:NSF_Phi}\\
		& = n(n+1)\int ds\int dt\,\rho_+(s)\,\rho_-(t) \nonumber\\
		& \quad \times \int_{0}^{1}dx\int_{0}^{1-x}dy \,(1-x-y)^{n-3} \nonumber\\
		& \quad \times \phi\left( \dfrac{\gamma+(x+y)P^2/4-xs-yt }{1-x-y}, \dfrac{z-x+y}{1-x-y} \right).\label{eq:NSF_Phi_reduced}
	\end{align}
	When the Nakanishi spectral function ${\Phi(\gamma,z)}$ is given by Eq.~\eqref{eq:NSF_Phi}, the momentum-space definition of the BSWF in Eq.~\eqref{eq:def_BSWF} is satisfied. Equation~\eqref{eq:NSF_Phi_reduced} is obtained as a simplification of Eq.~\eqref{eq:NSF_Phi} after eliminating the spectral variables $z'$ and $\gamma'$ utilizing properties of $\delta$-functions. 
	
	An alternative simplification that removes Feynman parameters $x$ and $y$ is preferred, since integrals with respect to the native variables of $\phi(\gamma,z)$ remain afterward. 
	This can be achieved with the geometric meaning of $\delta$-functions. Specifically the condition that arguments of $\delta$-functions vanish can be understood as straight lines in the $x$-$y$ plane. Points on these two lines respectively named $l_1$ and $l_2$ satisfy the following equations:
	\begin{subequations}\label{eq:def_l1_l2}
	\begin{align}
		& x-y+(1-x-y)z'-z=0, \\
		& xs+yt+(1-x-y)\gamma'-(x+y)P^2/4-\gamma=0. 
	\end{align}
	\end{subequations}
	The coordinates of their intersection point ${B=(x_0,y_0)}$ are given by
	\begin{subequations}
		\begin{align}
			& x_0 = [(1+z')\gamma-(1+z)\gamma'+(z-z')(t-P^2/4)]\,/\,J,\\
			& y_0 = [(1-z')\gamma-(1-z)\gamma'+(z'-z)(s-P^2/4)]\,/\,J,
		\end{align}
	where the signed Jacobian
	\begin{equation}
		J = s+t+(s-t)z'-2(\gamma'+P^2/4)
	\end{equation}
	is the determinant of the coefficient matrix when Eq.~\eqref{eq:def_l1_l2} is treated as a system of linear equations for $x$ and $y$. Let us also define ${z_0 = 1-x_0-y_0}$ such that
	\begin{equation}
		z_0 = {\left[s+t+(s-t)z-2\left(\gamma+P^2/4\right)\right]/J}. \label{eq:def_z_0}
	\end{equation}
	\end{subequations}
	When the point $B$ falls within the support of the Feynman parameters, the integrals will only sample this point. The integrals with respect $x$ and $y$ vanish otherwise. 
	
	\begin{figure}
		\centering
		\includegraphics[width=\linewidth]{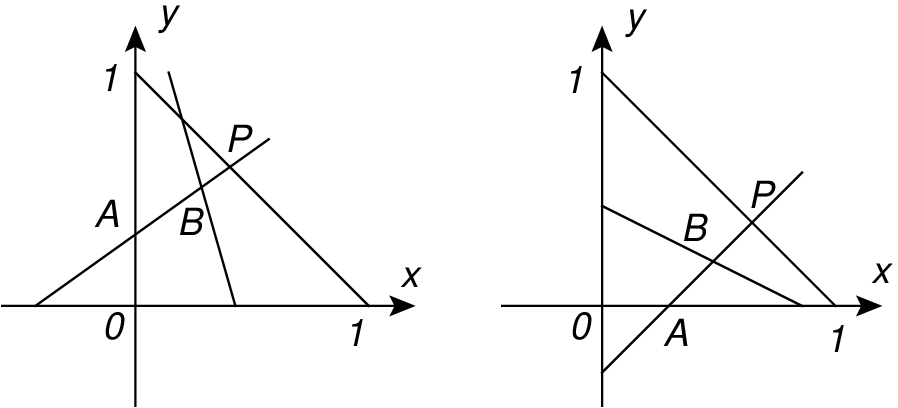}
		\caption{Illustrations of the Feynman-parameter space intersected by the straight lines corresponding to the arguments of $\delta$-functions in Eq.~\eqref{eq:NSF_Phi}. The left panel corresponds to the scenario where $z\leq z'$. While the right panel corresponds to $z\geq z'$.}
		\label{fig:bsemergefp}
	\end{figure}
	Because line $l_1$ can not be parallel to ${x+y=1}$, they must intersect at the point ${P=\left( (1+z)/2, (1-z)/2 \right)}$. Meanwhile, line $l_1$ intersects with the $y$-axis at point ${\left(0,(z'-z)/(1+z')\right)}$ and with the $x$-axis at a different point ${\left(-(z'-z)/(1-z'),0\right)}$. Expect for the special case of ${z=z'}$, these two points do not simultaneously locate within the support of the Feynman parameters. We therefore consider the following $2$ separate scenarios. 
	\begin{enumerate}
		\item Under the conditions that ${-1\leq z\leq z'\leq 1}$, the intersection of $l_1$ and the $y$-axis locates within the Feynman parameter space. As illustrated by the left panel of Fig.~\ref{fig:bsemergefp}, the integrals with respect to the Feynman parameters are nonzero only if the intersection of $l_1$ and $l_2$ falls on the line segment $\overline{AP}$ with ${A=\left(0,(z'-z)/(1+z')\right)}$. This is ensured when ${0\leq x_0\leq (1+z)/2}$ is satisfied, which indicates either for $J>0$
		\begin{subequations}
		\begin{equation}
			\begin{cases}
				2(\gamma+P^2/4) \leq s+t+(s-t)z \\
				(1+z)\gamma' \leq (1+z')\gamma-(z'-z)(t-P^2/4)
			\end{cases},
		\end{equation} 
		or for $J<0$
		\begin{equation}
			\begin{cases}
				2(\gamma+P^2/4) \geq s+t+(s-t)z \\
				(1+z)\gamma' \geq (1+z')\gamma-(z'-z)(t-P^2/4)
			\end{cases}.
		\end{equation}
		\end{subequations}
		\item While under the conditions that ${-1\leq z'\leq z\leq 1}$, the line $l_1$ intersection with the $x$-axis locates within the support of Feynman parameters. As shown on the right panel of Fig.~\ref{fig:bsemergefp}, the integrals with respect to the Feynman parameters are nonzero only if the intersection of $l_1$ and $l_2$ falls on the line segment $\overline{AP}$ with ${A=\left(-(z'-z)/(1-z'),0\right)}$. This is ensured when ${0\leq y_0\leq (1-z)/2}$, which indicates either for $J>0$
		\begin{subequations}
		\begin{equation}
			\begin{cases}
				2(\gamma+P^2/4) \leq s+t+(s-t)z \\
				(1-z)\gamma' \leq (1-z')\gamma-(z-z')(s-P^2/4)
			\end{cases},
		\end{equation} 
		or for $J<0$
		\begin{equation}
			\begin{cases}
				2(\gamma+P^2/4) \geq s+t+(s-t)z \\
				(1-z)\gamma' \geq (1-z')\gamma-(z-z')(s-P^2/4)
			\end{cases}.
		\end{equation}
		\end{subequations}
	\end{enumerate}
	The integrals with respect to the Feynman parameters are then reduced to $\theta$-functions of variables $\gamma,\,z,\,s,\,t,\,\gamma',$ and $z'$. For notational convenience, let us define the threshold function
	\begin{equation}
		\gamma_{\mathrm{th}}(z,\gamma,u,z') = \dfrac{1+z'}{1+z}\gamma - \dfrac{z'-z}{1+z}\left(u-\dfrac{P^2}{4} \right). \label{eq:def_fun_gamma_th}
	\end{equation}
	We subsequently obtain the following reduction of integrals in Eq.~\eqref{eq:NSF_Phi}:
	\begin{align*}
		& \int dz'\int d\gamma'\int_{0}^{1}dx\int_{0}^{1-x}dy\,(1-x-y)^{n-1}\nonumber\\
		&\quad \times \delta\left(x-y+(1-x-y)z'-z \right) \nonumber\\
		&\quad \times \delta\left(xs+yt+(1-x-y)\gamma'-(x+y)P^2/4-\gamma \right)\nonumber\\
		& = \left[ s+t+(s-t)z-2\left(\gamma+P^2/4\right) \right]^{n-1}  \nonumber\\
		& \quad \times \Bigg\{ \theta\left( [s+t+(s-t)z] - 2\left(\gamma+P^2/4 \right) \right) \nonumber\\
		& \times \left[ \int_{-1}^{z}dz' \int_{0}^{\gamma_{\mathrm{th}}(-z,\gamma,s,-z')}d\gamma'+ \int_{z}^{1}dz' \int_{0}^{\gamma_{\mathrm{th}}(z,\gamma,t,z')}d\gamma' \right] \nonumber\\
		& -\theta\left( 2\left( \gamma+P^2/4 \right) - [s+t+(s-t)z] \right) \nonumber\\
		& \times \left[ \int_{-1}^{z}dz'\int_{\gamma_{\mathrm{th}}(-z,\gamma,s,-z')}^{+\infty} d\gamma' + \int_{z}^{1}dz'\int_{\gamma_{\mathrm{th}}(z,\gamma,t,z')}^{+\infty} d\gamma' \right] \nonumber\\
		& \quad \Bigg\} \left[ s+t+(s-t)z'- 2\left(\gamma'+P^2/4 \right)\right]^{-n},
	\end{align*}
	where the index of the last factor is $-n$ instead of ${-n+1}$ due to the Jacobian in mapping the $\delta$-functions onto those of $x$ and $y$. Equation~\eqref{eq:NSF_Phi} then becomes
	\begin{align}
		& \Phi(\gamma,z) = n(n+1)\int ds\int dt \,\rho_+(s)\,\rho_-(t) \nonumber\\
		\times & \left[s+t+(s-t)z-2\left(\gamma+P^2/4\right)\right]^{n-1}\nonumber\\
		\times & \Bigg\{ \theta\left([s+t+(s-t)z]-2\left(\gamma+P^2/4 \right) \right) \nonumber\\
		\times & \left[\int_{-1}^{z}dz' \int_{0}^{\gamma_{\mathrm{th}}(-z,\gamma,s,-z')}d\gamma'+ \int_{z}^{1}dz' \int_{0}^{\gamma_{\mathrm{th}}(z,\gamma,t,z')}d\gamma'\right] \nonumber\\
		- & \theta\left(2\left(\gamma+P^2/4 \right)-[s+t+(s-t)z] \right) \nonumber\\
		\times & \left[\int_{-1}^{z}dz'\int_{\gamma_{\mathrm{th}}(-z,\gamma,s,-z')}^{+\infty} d\gamma' + \int_{z}^{1}dz'\int_{\gamma_{\mathrm{th}}(z,\gamma,t,z')}^{+\infty} d\gamma' \right] \Bigg\} \nonumber\\
		\times & \left[ s+t+(s-t)z'-2\left(\gamma'+P^2/4 \right)\right]^{-n}\,\phi(\gamma',z'),\label{eq:NIR_merge}
	\end{align}
	with thresholds of $\gamma'$ implied by $\phi(\gamma',z')$. The threshold $\Gamma_{\mathrm{th}}(z)$ for $\Phi(\gamma,z)$ is indicated by the $\theta$-functions in Eq.~\eqref{eq:NIR_merge}. In the case of $n=1$, Eq.~\eqref{eq:NIR_merge} is reduced to 
	\begin{align}
		& \Phi(\gamma,z) = 2\int ds\int dt \,\rho_+(s)\,\rho_-(t)\, \nonumber\\
		& \times \Bigg\{ \theta\left([s+t+(s-t)z]-2\left(\gamma+P^2/4 \right) \right) \nonumber\\
		& \times \left[\int_{-1}^{z}dz' \int_{0}^{\gamma_{\mathrm{th}}(-z,\gamma,s,-z')}d\gamma'+ \int_{z}^{1}dz' \int_{0}^{\gamma_{\mathrm{th}}(z,\gamma,t,z')}d\gamma'\right] \nonumber\\
		& -\theta\left(2\left(\gamma+P^2/4 \right)-[s+t+(s-t)z] \right) \nonumber\\
		& \times \left[\int_{-1}^{z}dz'\int_{\gamma_{\mathrm{th}}(-z,\gamma,s,-z')}^{+\infty} d\gamma' + \int_{z}^{1}dz'\int_{\gamma_{\mathrm{th}}(z,\gamma,t,z')}^{+\infty} d\gamma' \right] \Bigg\} \nonumber\\
		& \times \dfrac{\phi(\gamma',z')}{ s+t+(s-t)z'-2\left(\gamma'+P^2/4 \right)}\,. \label{eq:BEA_merge_n1} 
	\end{align}
	\subsection{BSE in Minkowski space\label{ss:bse_nakanishi}}
	\begin{figure}
		\centering
		\includegraphics[width=\linewidth]{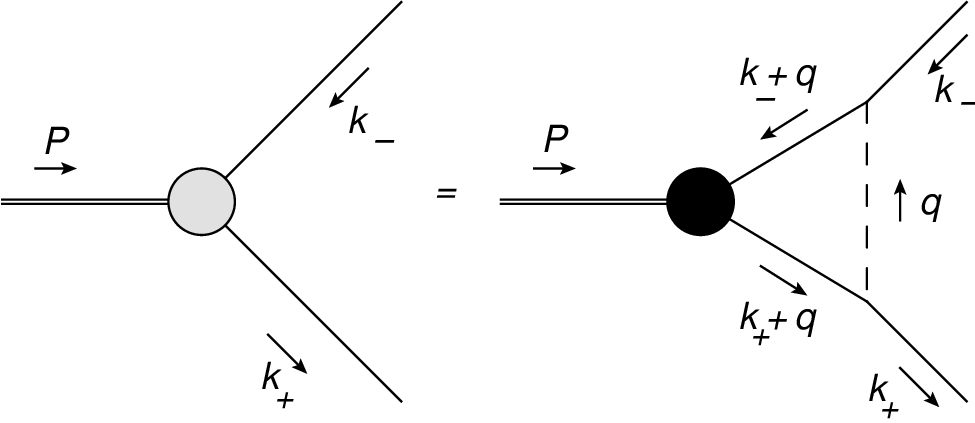}
		\caption{Diagrammatic representation of the BSE in the rainbow-ladder truncation.}
		\label{fig:nsrscalar}
	\end{figure}
	After combining propagators with BSA to form the BSWF, the BSE in the momentum space is given by 
	\begin{equation}
		\psi(k^2,k\cdot P) = ig^2\int d\underline{q}\,\chi((k+q)^2,(k+q)\cdot P)D_0(q),
	\end{equation}
	which is diagrammatically represented in Fig.~\ref{fig:nsrscalar}. Here $g$ is the coupling constant of the constituents with the exchange particle. In the massive variation of the Wick-Cutkosky model, the propagator of the exchange particle of mass $\mu$ is given by Eq.~\eqref{eq:def_bare_prop}. The integration measured $d\underline{q}$ is defined as
	\begin{equation}
		\int d\underline{q} = \dfrac{1}{(2\pi)^d}\int_{-\infty}^{+\infty}dq^0\prod_{j=1}^{d-1}\int_{-\infty}^{+\infty}dq^j.
	\end{equation}
	in $d$-dimensional spacetime. Substituting the Nakanishi representation for both the BSA and the BSWF into the momentum-space BSE gives
	\begin{align}
		& \!\!\!\! \int dz \int d\gamma \dfrac{\phi(\gamma,z)}{(k^2+zk\cdot P -\gamma+i\varepsilon )^n} = ig^2\int d\underline{q}\int dz \int d\gamma \nonumber\\
		& \times \dfrac{ \Phi(\gamma,z) }{ [(k+q)^2+z(k+q)\cdot P -\gamma]^{n+2} (q^2-\mu^2+i\varepsilon ) }. \label{eq:BSE_init_spectral_funs}
	\end{align}
	Here thresholds for the $\gamma$ variables are implied by $\phi(\gamma,z)$ and $\Phi(\gamma,z)$. The loop integral can be evaluated analytically using standard Feynman parametrization. After shifting $q^\mu \rightarrow q^\mu + k^\mu$, we have
	\begin{align}
		& ig^2\int d\underline{q}\dfrac{1}{ (q^2+zq\cdot P -\gamma+i\varepsilon)^{n+2} [(q-k)^2-\mu^2+i\varepsilon] } \nonumber\\
		& = ig^2(n+2) \int d\underline{l} \int_{0}^{1}dx\, \dfrac{x^{n+1}}{[l^2-\Delta]^{n+3}} \nonumber\\
		& = \dfrac{(-1)^ng^2}{(4\pi)^2} \dfrac{\Gamma(n+3-d/2)}{\Gamma(n+2)} \int_{0}^{1}dx\,x^{n+1} \dfrac{(4\pi)^{2-d/2}}{\Delta^{n+3-d/2}},
	\end{align}
	with 
	\begin{align*}
		& \Delta = x\gamma+(1-x)(\mu^2-k^2)+\left[xz\,P/2-(1-x)k \right]^2\nonumber\\
		&=-x(1-x)\bigg\{k^2+zk\cdot P -\left[\dfrac{\gamma}{1-x}+\dfrac{\mu^2}{x}+\dfrac{xz^2P^2}{4(1-x)} \right] \bigg\}
	\end{align*}
	and the number of spacetime dimensions given by ${d=4-2\epsilon}$. Although not explicitly required in the model we are considering, the parameter $\epsilon$ is used in dimensional regularization, which is not the same as the $\varepsilon$ in the Feynman prescription. When $d=4$ Eq.~\eqref{eq:BSE_init_spectral_funs} becomes 
	\begin{align}
		& \int dz \int d\gamma\, \dfrac{\phi(\gamma,z)}{(k^2+zk\cdot P -\gamma+i\varepsilon )^n}\nonumber\\
		& = \int dz \int d\gamma\, \Phi(\gamma,z) \dfrac{(-1)^ng^2}{(4\pi)^2} \dfrac{1}{n+1} \int_{0}^{1}dx\, \dfrac{x^{n+1}}{\Delta^{n+1}} \nonumber\\
		& = - \dfrac{g^2}{(4\pi)^2} \dfrac{1}{n+1} \int dz \int d\gamma\, \Phi(\gamma,z) \int_0^1 dx\, \dfrac{1}{(1-x)^{n+1}} \nonumber\\
		& \quad \times \bigg\{k^2+zk\cdot P -\left[\dfrac{\gamma}{1-x}+\dfrac{\mu^2}{x}+\dfrac{xz^2P^2}{4(1-x)} \right] \bigg\}^{-n-1}. \label{eq:BSE_loop_itg_spectral_funs}
	\end{align}
	Comparing the momentum dependence on both sides of Eq.~\eqref{eq:BSE_loop_itg_spectral_funs} gives 
	\begin{align}
		& \phi(\gamma,z) = -\dfrac{g^2}{(4\pi)^2} \dfrac{1}{n(n+1)} \int dz'\int d\gamma' \int_{0}^{1}dx \dfrac{\Phi(\gamma',z')}{(1-x)^{n+1}}\nonumber\\
		& \times \delta(z-z')\,\delta\left( \gamma - \left[ \dfrac{\gamma'}{1-x} + \dfrac{\mu^2}{x} + \dfrac{xz'^2P^2}{4(1-x)} \right] \right) \dfrac{\partial}{\partial\gamma}\label{eq:BSE_phi_Phi_ori}\\
		& = -\dfrac{g^2}{(4\pi)^2} \dfrac{1}{n(n+1)} \int_{0}^{1}dx \dfrac{1}{(1-x)^n} \nonumber\\
		& \quad \times \Phi\left((1-x)(\gamma -\mu^2/x)-xz^2P^2/4,\,z \right)\dfrac{\partial}{\partial \gamma }. \label{eq:BSE_phi_Phi_xitg}
	\end{align}
	The factor of ${\partial/\partial \gamma}$ is due to the mismatched powers of momentum in BSE, which is accepted since we allow $\phi(\gamma,z)$ to be a distribution. This mismatch happens because the coupling constant carries the dimension of momentum. At both limits of $x\rightarrow 0$ and $x\rightarrow 1$, the integral is finite as ${\Phi(\gamma,z)}$ vanishes when ${\gamma<0}$. 
	
	Similar to Subsection~\ref{ss:BSWF_merge}, a preferred alternative expression of $\psi(\gamma,z)$ in terms of $\Phi(\gamma,z)$ can be obtained by eliminating the integral with respect to the Feynman parameter $x$. Specifically based on
	\begin{align}
		& \delta\left( \gamma - \left[ \dfrac{\gamma'}{1-x} + \dfrac{\mu^2}{x} + \dfrac{xz'^2P^2}{4(1-x)} \right] \right) = x(1-x) \nonumber\\
		& \times \delta\left( ( \gamma+z'^2 P^2/4 ) x^2+(\gamma'-\mu^2-\gamma)x+\mu^2 \right),\label{eq:delta_x_BSE}
	\end{align}
	we have the following $2$ roots for $x$:
	\begin{align}
		& x_\pm(\gamma',z',\gamma) = \nonumber\\
		& \dfrac{ (\gamma+\mu^2-\gamma') \pm \sqrt{ (\gamma+\mu^2-\gamma')^2-4\mu^2(\gamma+z'^2P^2/4)} }{ 2(\gamma+z'^2P^2/4) }.
	\end{align}
	Notice that on both ends $x=0$ and $x=1$ the argument of the $\delta$-function on the right-hand side of Eq.~\eqref{eq:delta_x_BSE} as a quadratic function of $x$ is positive. These solutions are then real and fall into ${[0,1]}$ only if 
	\begin{subequations}\label{eq:itg_support_gammap}
	\begin{equation}
		\mu^2-\gamma-z'^2 P^2/2\leq \gamma' \leq \gamma+\mu^2-2\mu\sqrt{\gamma+z'^2P^2/4}.\label{eq:itg_support_gammap_main}
	\end{equation}
	The first condition comes from ${0\leq (x_++x_-)/2\leq 1}$. While the second one is a consequence of ${(\gamma+\mu^2-\gamma')^2}-{4\mu^2(\gamma+z'^2P^2/4)} \geq 0$. Equation~\eqref{eq:itg_support_gammap} further suggests 
	\begin{equation}
		\gamma +z'^2P^2/4 \geq \mu^2. \label{eq:itg_support_gammap_2ed}
	\end{equation}
	\end{subequations}
	When Eq.~\eqref{eq:itg_support_gammap} is satisfied, Eq.~\eqref{eq:BSE_phi_Phi_ori} becomes
	\begin{align}
		& \phi(\gamma,z) = -\dfrac{g^2}{(4\pi)^2} \dfrac{1}{n(n+1)} \int dz' \int d\gamma' \int_{0}^{1}dx\, \Phi(\gamma',z') \nonumber\\
		& \dfrac{x\, \delta(z-z')}{(1-x)^n} \dfrac{ \delta\left(x-x_-(\gamma',z',\gamma) \right) + \delta\left(x-x_+(\gamma',z',\gamma)\right) }{ \vert \gamma+z^2 P^2/4\vert [x_+(\gamma',z',\gamma)-x_-(\gamma',z',\gamma)] }\dfrac{\partial}{\partial\gamma}\nonumber\\
		& = -\dfrac{g^2}{(4\pi)^2} \dfrac{1}{n(n+1)} \int_{\mu^2-\gamma-z^2P^2/2}^{\gamma+\mu^2-2\mu\sqrt{\gamma+z^2P^2/4}} d\gamma' \nonumber\\
		& \quad \times \dfrac{ \Phi(\gamma',z) }{ \sqrt{ (\gamma+\mu^2-\gamma')^2-4\mu^2(\gamma+z^2P^2/4) } }\nonumber\\
		& \quad \bigg\{ \dfrac{x_-(\gamma',z,\gamma)}{[1-x_-(\gamma',z,\gamma) ]^n} + \dfrac{x_+(\gamma',z,\gamma)}{[1-x_+(\gamma',z,\gamma) ]^n} \bigg\} \dfrac{\partial}{\partial\gamma}, 
	\end{align}
	as an alternative to Eq.~\eqref{eq:BSE_phi_Phi_xitg} that keeps the integral with respect to the spectral variables. Specifically in the case of $n=1$, we obtain
	\begin{align}
		& \phi(\gamma,z) = -\dfrac{g^2}{(4\pi)^2}\int_{\mu^2-\gamma-z^2P^2/2}^{\gamma+\mu^2-2\mu\sqrt{\gamma+z^2P^2/4}} d\gamma'\dfrac{\gamma-\mu^2-\gamma'}{2(\gamma'+z^2P^2/4)} \nonumber\\
		&\quad \times \dfrac{\Phi(\gamma',z)}{\sqrt{(\gamma+\mu^2-\gamma')^2-4\mu^2(\gamma+z^2P^2/4)}}\dfrac{\partial}{\partial\gamma}. \label{eq:BSE_phi_Phi_n1}
	\end{align}
	The integral with respect to $\gamma'$ can be simplified with the application of the variable transform
	\begin{equation}
		x=\mathrm{arccosh}\dfrac{\gamma+\mu^2-\gamma'}{2\mu\sqrt{\gamma+z^2P^2/4}}
	\end{equation}
	such that $\gamma'=\gamma+\mu^2-2\mu\sqrt{\gamma+z^2P^2/4}\cosh x$. Applying this variable transform also removes the inverse square-root singularity at one end of the integration limit. The BSE for the Nakanishi spectral functions in the form of Eq.~\eqref{eq:BSE_phi_Phi_n1} then becomes
	\begin{align}
		& \phi(\gamma,z) = - \dfrac{g^2}{(4\pi)^2}\int_{0}^{\xi(\gamma,z)}dx \big(\mu\sqrt{\gamma+z^2P^2/4}\cosh x-\mu^2\big)\nonumber\\
		& \times \dfrac{\Phi(\gamma+\mu^2-2\mu\sqrt{\gamma+z^2P^2/4}\cosh x,\,z )}{\gamma+\mu^2+z^2P^2/4-2\mu\sqrt{\gamma+z^2P^2/4}\cosh x} \dfrac{\partial}{\partial \gamma},\label{eq:BES_vari_trans}
	\end{align}
	where the upper limit of integration is given by
	\begin{equation}
		\cosh \xi(\gamma,z) = \min \left( \dfrac{\sqrt{\gamma+z^2P^2/4}}{\mu},\, \dfrac{\gamma+\mu^2-\Gamma_{\mathrm{th}}(z)}{2\mu\sqrt{\gamma+z^2P^2/4}} \right)
	\end{equation}
	with $\Gamma_{\mathrm{th}}(z)$ being the threshold above which $\Phi(\gamma,z)$ is nonzero. By requesting the upper integration limit of Eq.~\eqref{eq:BSE_phi_Phi_n1} to be greater than $\Gamma_{\mathrm{th}}(z)$, we find the $\gamma$-threshold of $\phi(\gamma,z)$ as
	\begin{equation}
		g_{\mathrm{th}}(z) = \Gamma_{\mathrm{th}}(z)+\mu^2+2\mu\sqrt{ \Gamma_{\mathrm{th}}(z) +z^2P^2/4},
		\label{eq:gamma_threshold_phi}
	\end{equation}
	Notice that the condition
	\[\Gamma_{\mathrm{th}}(z)+\mu^2+2\mu\sqrt{\Gamma_{\mathrm{th}}(z)+z^2P^2/4 }\geq \mu^2-z^2P^2/4 \]
	holds true, which ensures Eq.~\eqref{eq:itg_support_gammap_2ed} because the BSE does not operate on the $z$ variable. 
	
	For numerical solutions of the BSE in Eq.~\eqref{eq:BES_vari_trans}, let us define an auxiliary function $\Theta(\gamma,z)$ by factoring the derivative operator in $\phi(\gamma,z)$ through
	\begin{equation}
		\phi(\gamma,z)\equiv- \Theta(\gamma,z)\,\dfrac{\partial}{\partial \gamma}. \label{eq:def_Theta_gamma_z}
	\end{equation}
	Equation~\eqref{eq:BES_vari_trans} subsequently becomes 
	\begin{align}
		& \Theta(\gamma,z) = \dfrac{g^2}{(4\pi)^2}\int_{0}^{\xi(\gamma,z)}dx\, (\mu\sqrt{\gamma+z^2P^2/4}\cosh x-\mu^2)\nonumber\\
		& \times \dfrac{\Phi( \gamma+\mu^2-2\mu\sqrt{\gamma+z^2P^2/4}\cosh x, z ) }{\gamma+\mu^2+z^2P^2/4-2\mu\sqrt{\gamma+z^2P^2/4}\cosh x}. \label{eq:BSE_vari_trans_Theta}
	\end{align}

	Because both $\Phi(\gamma,z)$ and $\Theta(\gamma,z)$ are distributions, the direct implementation of the partial derivative with respect to $\gamma$ from Eq.~\eqref{eq:def_Theta_gamma_z} is not valid for Eq.~\eqref{eq:BEA_merge_n1}. Instead in order to maintain Eq.~\eqref{eq:def_BSWF} in the momentum space, we need to develop the corresponding distribution identity. Assuming integration-by-part is valid for distributions, we obtain the following identity
	\begin{align}
		& \int_{\gamma_{\mathrm{th}}}^{+\infty}d\gamma'\dfrac{1}{\gamma'-m^2+P^2/4}\dfrac{\partial}{\partial\gamma'}\delta(\gamma'-\gamma_0)\nonumber\\
		& = \lim\limits_{\gamma'\rightarrow +\infty}\dfrac{\delta(\gamma'-\gamma_0)}{\gamma'-m^2+P^2/4}-\lim\limits_{\gamma'\rightarrow \gamma_{\mathrm{th}}}\dfrac{\delta(\gamma'-\gamma_0)}{\gamma'-m^2+P^2/4} \nonumber\\
		& \quad +\int_{\gamma_{\mathrm{th}}}^{+\infty}d\gamma'\dfrac{\delta(\gamma'-\gamma_0)}{(\gamma'-m^2+P^2/4)^2}.\label{eq:derivative_phi_merge_bare_propagator_n1}
	\end{align}
	In the case where $\gamma_0$ is finite, the first term in Eq.~\eqref{eq:derivative_phi_merge_bare_propagator_n1} vanishes. The second term is proportional to a $\delta$-function when ${\gamma_{\mathrm{th}}=\gamma_0}$. While the third term contains a factor of $\theta$-function that reflects the condition of ${\gamma_0>\gamma_{\mathrm{th}}}$. 
	
	When the propagators of the constituents are bare, we have the threshold of $\Phi(\gamma,z)$ given by 
	\begin{equation}
		\Gamma_{\mathrm{th}}(z) = m^2 - P^2/4. \label{eq:def_Gamma_th_bare_prop}
	\end{equation}
	The merging condition in Eq.~\eqref{eq:BEA_merge_n1} becomes 
	\begin{align}
		& \Phi(\gamma,z)
		= \theta(\gamma-m^2+P^2/4) \bigg\{ \int_{-1}^{z}dz'\int_{\gamma_{\mathrm{th}}(-z,\gamma,s,-z')}^{+\infty} d\gamma' \nonumber\\
		& \quad + \int_{z}^{1}dz'\int_{\gamma_{\mathrm{th}}(z,\gamma,t,z')}^{+\infty} d\gamma' \bigg\} \dfrac{\phi(\gamma',z')}{\gamma'-m^2+P^2/4} . \label{eq:BEA_merge_n1_rdd} 
	\end{align}
	In order to apply Eq.~\eqref{eq:def_Theta_gamma_z}, we then need to identify the second $\delta$-function term on the right-hand side of Eq.~\eqref{eq:derivative_phi_merge_bare_propagator_n1}. The following limit
	\begin{align}
		& \lim\limits_{\gamma'\rightarrow\gamma_{\mathrm{th}}(\pm z,\gamma,m^2,\pm z_0)}\dfrac{\delta(\gamma'-\gamma_0)}{\gamma'-m^2+P^2/4} \nonumber\\
		& =
		\begin{cases}
			0\quad(\mathrm{for}~\gamma_{\mathrm{th}}(\pm z,\gamma,m^2,\pm z_0)\neq \gamma_0)\\
			\infty\quad(\mathrm{for}~\gamma_{\mathrm{th}}(\pm z,\gamma,m^2,\pm z_0)=\gamma_0) 
		\end{cases}
	\end{align}
	can be recognized as a $\delta$-function of $\gamma$. Applying the definition of ${\gamma_{\mathrm{th}}(z,\gamma,u,z',P^2)}$ in Eq.~\eqref{eq:def_fun_gamma_th}, we could demonstrate that
	\begin{align}
		& \int d\gamma' \lim\limits_{\gamma'\rightarrow\gamma_{\mathrm{th}}(\pm z,\gamma,m^2,\pm z_0)}\dfrac{\delta(\gamma'-\gamma_0)}{\gamma'-m^2+P^2/4} = \nonumber\\
		& \dfrac{1\pm z}{(1\pm z_0)(\gamma_0-m^2+P^2/4)} \times \nonumber\\
		& \delta\left(\gamma-\left[\dfrac{1\pm z}{1\pm z_0}\left(\gamma_0-m^2+\dfrac{P^2}{4} \right) +m^2-\dfrac{P^2}{4} \right] \right).
	\end{align}
	Assuming that $\Theta(\gamma,z)$ is a regular function in Eq.~\eqref{eq:def_Theta_gamma_z}, let us consider
	\begin{equation}
		\phi(\gamma,z)=-\delta(\gamma-\gamma_0)\,\delta(z-z_0)\dfrac{\partial}{\partial\gamma} \label{eq:phi_gamma_z_single_elem}
	\end{equation}
	to isolate the contribution to Eq.~\eqref{eq:NIR_merge} from a single element of $\Theta(\gamma,z)$. After applying Eq.~\eqref{eq:derivative_phi_merge_bare_propagator_n1} within this condition converts Eq.~\eqref{eq:BEA_merge_n1_rdd} to
	\begin{align}
		& \Phi(\gamma,z) = -\theta( \gamma-m^2+P^2/4 ) \bigg[ \int_{-1}^{z}dz' \int_{\gamma_{\mathrm{th}}(-z,\gamma,s,-z')}^{+\infty} d\gamma' \nonumber\\
		& \quad + \int_{z}^{1}dz' \int_{\gamma_{\mathrm{th}}(z,\gamma,t,z')}^{+\infty} d\gamma' \bigg] \dfrac{\delta(\gamma'-\gamma_0)\,\delta(z'-z_0)}{\gamma'-m^2+P^2/4} \dfrac{\partial}{\partial\gamma'} \nonumber\\
		& = \theta(\gamma-m^2+P^2/4) \,\Bigg\{ \theta(z-z_0) \,\bigg\{ \nonumber\\
		& \quad -\lim\limits_{\gamma'\rightarrow \gamma_{\mathrm{th}}(-z,\gamma,m^2,-z_0)} \dfrac{\delta(\gamma'-\gamma_0)}{\gamma'-m^2+P^2/4} \nonumber\\
		& \quad +\int_{\gamma_{\mathrm{th}}(-z,\gamma,m^2,-z_0)}^{+\infty}d\gamma'\dfrac{\delta(\gamma'-\gamma_0)}{(\gamma'-m^2+P^2/4)^2} \bigg\}\nonumber
	\end{align}
	\begin{align}
		& \quad + \theta(z_0-z)\,\bigg\{-\lim\limits_{\gamma'\rightarrow \gamma_{\mathrm{th}}(z,\gamma,m^2,z_0)}\dfrac{\delta(\gamma'-\gamma_0)}{\gamma'-m^2+P^2/4} \nonumber\\
		& \quad +\int_{\gamma_{\mathrm{th}}(z,\gamma,m^2,z_0)}^{+\infty}d\gamma'\dfrac{\delta(\gamma'-\gamma_0)}{(\gamma'-m^2+P^2/4)^2} \bigg\} \Bigg\},
	\end{align}
	which gives the spectral function $\Phi(\gamma,z)$ when $\phi(\gamma,z)$ is given by Eq.~\eqref{eq:phi_gamma_z_single_elem}. In general when $\phi(\gamma,z)$ is given by Eq.~\eqref{eq:def_Theta_gamma_z}, we have
	\begin{align}
		& \Phi(\gamma,z) = \theta(\gamma-m^2+P^2/4) \int dz' \int d\gamma'\,\Theta(\gamma',z')\, \Bigg\{ \nonumber\\
		& \theta(z-z') \, \bigg\{ -\dfrac{1- z}{(1-z')(\gamma'-m^2+P^2/4)} \nonumber\\
		& \quad \times \delta\left(\gamma-\left[\dfrac{1- z}{1- z'}\left(\gamma'-m^2+\dfrac{P^2}{4} \right) +m^2-\dfrac{P^2}{4} \right] \right)\nonumber\\
		& \quad +\dfrac{1}{(\gamma'-m^2+P^2/4)^2} \nonumber\\
		& \quad \times \theta \left(\left[\dfrac{1- z}{1- z'}\left(\gamma'-m^2+\dfrac{P^2}{4} \right) +m^2-\dfrac{P^2}{4} \right]-\gamma \right) \bigg\} \nonumber\\
		& + \theta(z'-z)\bigg\{-\dfrac{1+ z}{(1+ z')(\gamma'-m^2+P^2/4)} \nonumber\\
		& \quad \times \delta\left(\gamma-\left[\dfrac{1+ z}{1+ z'}\left(\gamma'-m^2+\dfrac{P^2}{4} \right) +m^2-\dfrac{P^2}{4} \right] \right)\nonumber\\
		& \quad +\dfrac{1}{(\gamma'-m^2+P^2/4)^2} \nonumber\\
		& \quad \times \theta\left(\left[\dfrac{1+ z}{1+ z'}\left(\gamma'-m^2+\dfrac{P^2}{4} \right) +m^2-\dfrac{P^2}{4} \right] -\gamma\right) \bigg\} \Bigg\}. \label{eq:phi_gamma_z_Theta}
	\end{align}
	For a given combination of $\gamma$, $z$, and $z'$, the $\theta$-functions in the $5$-th line and the $9$-th line of Eq.~\eqref{eq:phi_gamma_z_Theta} generate the lower limit of integration with respect to $\gamma'$. Because the $\delta$-functions correspond to the sampling of $\Theta(\gamma',z')$ at the same limit, they are handled analytically in the computation of the $\gamma'$ integral. Together with the BSE in the form of Eq.~\eqref{eq:BSE_vari_trans_Theta}, we have arrived a closed system of equations that form the basis of an iterative solver for the Nakanishi spectral functions from the BSE. 
	\subsection{The normalization of the Bethe-Salpeter amplitude}
	With the rainbow-ladder kernel, the Cutkosky-Leon normalization in terms of the BSWF is given by
	\begin{subequations}
	\begin{equation}
		\int d\underline{k}\,\overline{\chi}(k,P)\left[\frac{\partial}{\partial P_\mu}D^{-1}_+\left(k_+ \right) D^{-1}_-\left(k_- \right) \right] 
		\chi(k,P) = 2iP^\mu,
	\end{equation}
	with $\overline{\chi}(k,P)$ being the conjugate BSWF~\cite{Kusaka:1997xd}. We then apply the definition of the BSWF in Eq.~\eqref{eq:def_BSWF} to obtain 
	\begin{equation}
		\int d\underline{k}\,\overline{\psi}(k,P)\left[\frac{\partial}{\partial P_\mu}D_+\left(k_+ \right)D_-\left(k_- \right) \right] \psi(k,P) = -2iP^\mu
	\end{equation}
	\end{subequations}
	as the normalization condition for the BSA. Applying the chain rule of the derivative gives
	\begin{equation}
		\int d\underline{k}\,\left[\overline{\psi}(k,P)\dfrac{\partial \chi(k,P)}{\partial P_\mu} -\overline{\chi}(k,P) \dfrac{\partial \psi(k,P)}{\partial P_\mu} \right] = -2iP^\mu,\label{eq:normalization_cond}
	\end{equation}
	which does not explicitly contain the propagators. Substituting the Nakanishi integral representations of the BSA and the BSWF in Eqs.~\eqref{eq:NIR_amplitude} and \eqref{eq:NIR_chi} into Eq.~\eqref{eq:normalization_cond} gives the normalization condition in terms of spectral functions solved from the BSE. 
	
	As suggested in Ref.~\cite{Kusaka:1997xd}, the physical normalization condition given by Eq.~\eqref{eq:normalization_cond} requires multidimensional integrals, which can be computationally demanding when compared to other operations in an iterative solver of the BSE. We therefore adopt an alternative normalization 
	\begin{equation}
		1 = \int_{-1}^{1}dz\int_{g_{\mathrm{th}}(z)}^{+\infty} d\gamma\, \dfrac{\Theta(\gamma,z)}{\gamma^\alpha},\label{eq:normalization_cond_lambda}
	\end{equation}
	with ${\alpha=2}$. Ensuring Eq.~\eqref{eq:normalization_cond_lambda} results in a rescaling factor of the function ${\Theta(\gamma,z)}$ at each step of iteration, the stable value of which is recognized as the eigenvalue $\lambda$ of the BSE. 
	\section{Numerical solutions of Minkowski-space BSE\label{sc:numeric}}
	\subsection{Radial variable transform and algorithm of adaptive grid for integral equations\label{ss:algorithm}}
	\begin{figure}
		\centering
		\includegraphics[width=\linewidth]{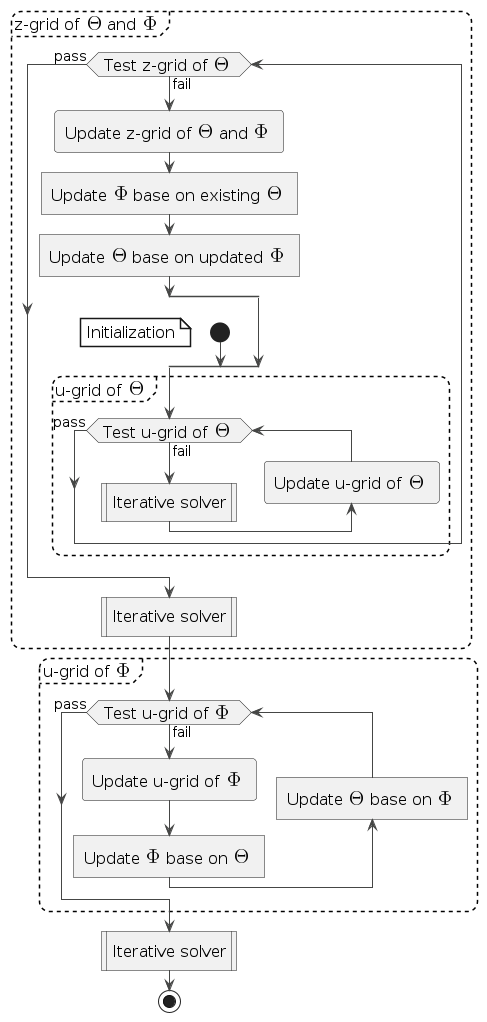}
		\caption{Activity diagram for the algorithm of adaptive grid in solving for the Nakanishi spectral functions from the BSE. }
		\label{fig:scalaradaptivegridbse}
	\end{figure}
	\begin{figure*}
		\centering
		\includegraphics[width=\linewidth]{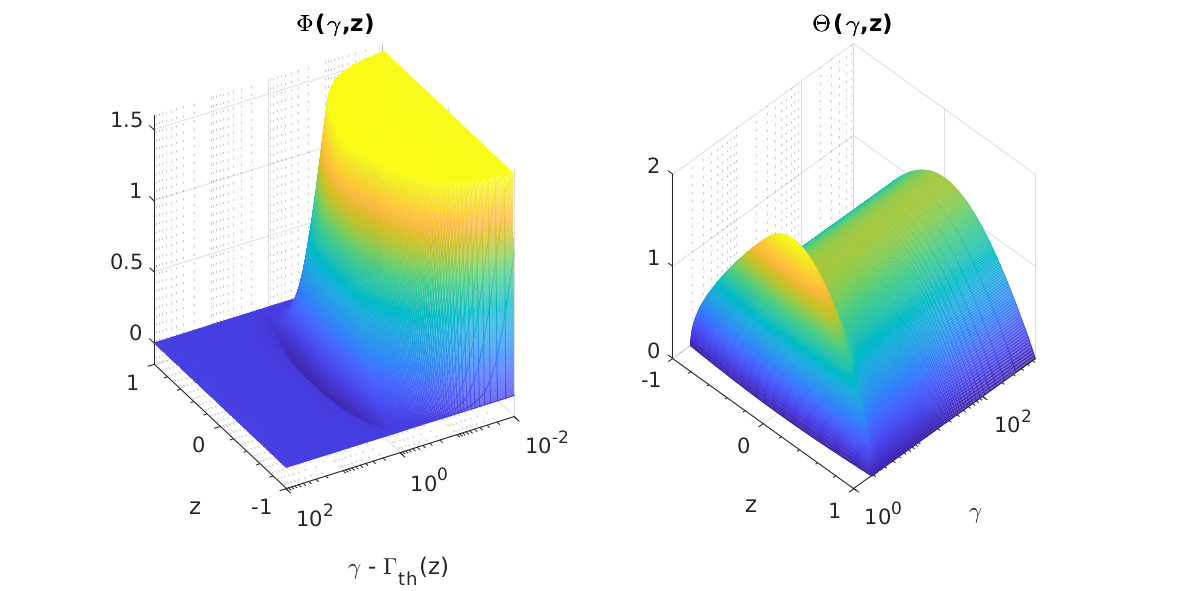}
		\caption{Numerical solutions of the Nakanishi spectral functions from the BSE. Left: spectral function $\Phi(\gamma,z)$ for the BSWF. Right: $\Theta(\gamma,z)$ as the nonderivative part of the spectral function for the BSA.}
		\label{fig:specfuns}
	\end{figure*}
	We would like to solve the Nakanishi spectral functions numerically from their BSE. Specifically, we work in the massive variation of the Wick-Cutkosky model where the propagators for the constituents and the exchange particle are all bare. We then choose the constituents to be identical particles of mass $m$. The spectral functions for their propagators are given by 
	\begin{equation}
		\rho_{\pm}(s) = \delta( s - m^2 ). 
	\end{equation}
	While the mass of the exchange particle is $\mu$. The depth of the bound is specified by a parameter $\eta$ such that $P^2$ as the square of bound state mass is related to the constituent mass through
	\begin{equation}
		P^2 = 4\eta^2\, m^2. \label{eq:def_eta_bound}
	\end{equation}
	For a given set of mass as input parameters, the coupling constant $g$ becomes an output from the iterative solver. 
	
	Because the upper limits of the integrals for the spectral variables are infinity, for a given value of angular variable $z$, we introduce a transformation for the radial variable $\gamma$ such that
	\begin{subequations}
	\begin{equation}
		u(\gamma,z)=\dfrac{\gamma-\gamma_{\mathrm{th}}(z)}{\gamma-\gamma_{\mathrm{th}}(z)+C(z)},\label{eq:def_vars_trans_u_of_gamma}
	\end{equation}
	with $\gamma_{\mathrm{th}}(z)$ being the threshold of the radial integral and $C(z)$ the scale of the transform. The variable $u$ then falls within ${[0,1]}$, suitable for numerical integration. The inverse transform is
	\begin{equation}
		\gamma(u,z)= C(z)\,u/ (1-u) + \gamma_{\mathrm{th}}(z),\label{eq:vars_trans_gamma_of_u}
	\end{equation}
	resulting in the following factor of integration measure:
	\begin{equation}
		d\gamma = C(z)\,/(1-u)^2\,du. \label{eq:measure_trans_gamma_of_u}
	\end{equation}
	\end{subequations}
	Because both $\gamma_{\mathrm{th}}(z)$ and $C(z)$ are allowed to depend on the variable $z$, the differential with respect to $\gamma$ in Eq.~\eqref{eq:measure_trans_gamma_of_u} is taken at fixed $z$. For a given $z$, the integral with respect to $\gamma$ from $\gamma_{\mathrm{th}}(z)$ to $+\infty$ then becomes 
	\begin{equation*}
		\int_{\gamma_{\mathrm{th}}(z)}^{+\infty}d\gamma=\int_{0}^{1}\dfrac{C(z)}{(1-u)^2}du. 
	\end{equation*}

	After defining the working variables, we then propose an algorithm to solve for the spectral functions from the BSE on an adaptive quadrature grid. This algorithm is illustrated in Fig.~\ref{fig:scalaradaptivegridbse}. 
	Because the BSE we have derived only involves operations of the radial variables, it is convenient to choose an identical grid for the $z$ variables in both $\Theta(\gamma,z)$ and $\Phi(\gamma,z)$. This choice avoids interpolation in the angular direction when solving the BSE. 
	
	At the initial step in solving the integral equations, we choose a reasonably sized initial grid for the angular and radial spectral variables. After making an initial guess of $\Phi(\gamma,z)$, the BSE is solved numerically with an iterative solver for the ground state. Specifically in each step of the iteration, the BSE in Eq.~\eqref{eq:BSE_vari_trans_Theta} is first used to compute $\Theta(\gamma,z)$ from a given $\Phi(\gamma,z)$. The merge identity in Eq.~\eqref{eq:phi_gamma_z_Theta} is then applied to update the $\Phi(\gamma,z)$ from the $\Theta(\gamma,z)$ obtained. The iterative solver stops when the desired tolerance conditions are met for changes in the function $\Theta(\gamma,z)$. 
	
	We apply the adaptive Simpson's rule for both the angular variable $z$ and the mapped radial variable $u$ in both spectral functions $\Theta(\gamma,z)$ and $\Phi(\gamma,z)$. The corresponding quadrature rule with tolerance estimate is given in the Appendix. The condition in Eq.~\eqref{eq:quad_error} to densify the grid requires the evaluation of a fixed integration. Specifically we choose the right-hand side of Eq.~\eqref{eq:normalization_cond_lambda} as the condition for the $z$-grid of function $\Theta(\gamma,z)$, as it ensures the adaptive grid is constructed to produce accurate eigenvalues. The corresponding condition for the $u$-grid of $\Theta(\gamma,z)$ at a given value of $z$ is subsequently based on
	\begin{equation}
		C(z)\,\int_{0}^{1}du\,\dfrac{\Theta(\gamma(u,z),z)}{[C(z)\,u + g_{\mathrm{th}}(z)\,(1-u)]^2}.\label{eq:quad_cond_u_Theta}
	\end{equation}
	For a given $z$-grid, we update the $u$-grid for the function $\Theta(\gamma,z)$ based on this condition. With each update, because the grid for the function $\Phi(\gamma,z)$ is unchanged, one step of the iterative solver is applied to update the solution of the BSE. 
	
	We then update the $z$-grid based on the adaptive algorithm. A corresponding $u$-grid of initial size is associated with each added point in $z$. The function $\Phi(\gamma,z)$ is then updated based on the previous version of the function $\Theta(\gamma,z)$,
	applying the merge identity in Eq.~\eqref{eq:phi_gamma_z_Theta}. The BSE in Eq.~\eqref{eq:BSE_vari_trans_Theta} is utilized next to update $\Theta(\gamma,z)$ based on the updated $\Phi(\gamma,z)$. The $z$-grid update is complete when the criteria do not require further updates in either the $z$-grid or the $u$-grid of the function ${\Theta(\gamma,z)}$. After updating the $z$-grid, there will be no more changes in the mesh grid of ${\Theta(\gamma,z)}$. The iterative solver is therefore called to obtain the solution of the BSE with current grid. 
	
	To reach the final grid for $\Phi(\gamma,z)$, the adaptive algorithm is called to update its $u$-grid. With each update, the merge identity in Eq.~\eqref{eq:phi_gamma_z_Theta} is applied to compute the new $\Phi(\gamma,z)$ based on the current $\Theta(\gamma,z)$. The function $\Theta(\gamma,z)$ is subsequently updated applying the BSE in Eq.~\eqref{eq:BSE_vari_trans_Theta} based on the updated $\Phi(\gamma,z)$. The reference value to densify the $u$-grid of $\Phi(\gamma,z)$ for a given $z$ is given by
	\begin{equation}
		C(z)\,\int_{0}^{1}du\,\dfrac{\Phi(\gamma(u,z),z)}{[C(z)\,u + \Gamma_{\mathrm{th}}(z)\,(1-u)]^2},
	\end{equation}
	as an analogue of Eq.~\eqref{eq:quad_cond_u_Theta}. When no further $u$-grid update is needed for the function $\Phi(\gamma,z)$, the iterative solver is called to get the final solution. 
	\subsection{Numerical solution of the BSE}
	Applying the algorithm discussed in the previous subsection, we proceed to find the numerical solutions of the BSE in terms of Nakanishi spectral functions $\Phi(\gamma,z)$ and $\Theta(\gamma,z)$. Specifically we first choose the mass of the constituents as the unit of measure for momenta, resulting in $m^2=1$. The exchange-particle mass is set to be half of the constituent mass $\mu = m / 2$. We then select $\eta = 0.6$ so that the mass of the bound state is $P^2 = 1.44$. With this set of model parameters, we obtain an eigenvalue of ${g^2/(4\pi)^2 = 1.940\,614}$ in units of the constituent mass, which is in agreement with Ref.~\cite{Kusaka:1997xd}. 
	
	The relative and absolute tolerances of the iterative solver are both $1.0\times 10^{-6}$ for the function $\Theta(\gamma,z)$. The iterative solver converges when either the relative change or the absolute change in $\Theta(\gamma,z)$ is below this tolerance for all grid points. The tolerances for conditions in adaptive-grid quadrature are $\epsilon_{\mathrm{tol}} = 1.0\times 10^{-6}$ for both the $u$ variable and the $z$ variable. The scale of the radial variable transformation is $C(z) = 5.0$ for both spectral functions. The number of quadrature points in the initial grid for the $z$ variable is based on $N_{\mathrm{init}} = 5$ and Eq.~\eqref{eq:def_N_init}. While for the $u$ variables, we choose $N_{\mathrm{init}} = 5$ for $\Phi(\gamma,z)$ and $N_{\mathrm{init}} = 0$ for $\Theta(\gamma,z)$. The resulting Nakanishi spectral functions from the iterative solver of the BSE with $\eta=0.6$ applying the adaptive-grid algorithm are illustrated in Fig.~\ref{fig:specfuns}. After adjusting all numerical tolerances to $1.0\times 10^{-4}$, we then obtain the dependence of the eigenvalue on the depth of bound in Table~\ref{tab:lambda_eta}. This result is again in agreement with Ref.~\cite{Kusaka:1997xd}.
	\begin{table}[h]
		\caption{The dependence of the coupling constant $\lambda = {g^2/(4\pi)^2}$ on the depth of the bound $\eta$ for the ground state. The second row corresponds to the result from Ref.~\cite{Kusaka:1997xd}, while our result is given in the third row.\label{tab:lambda_eta}}
		\begin{ruledtabular}
			\begin{tabular}{c|ccccccccc}
				$\eta$ & $0.0$ & $0.2$ & $0.4$ & $0.6$ & $0.8$ & $0.9$ & $0.99$ & $0.999$ \\
				\hline
				\cite{Kusaka:1997xd} & $2.5662$& $2.4988$ & $2.2937$ & $1.9402$ & $1.4056$ & $1.0350$ & $0.5168$ & $0.3853$ \\
				$\lambda$ & $2.5672$ & $2.4986$ & $2.2937$ & $1.9406$ & $1.4057$ & $1.0357$ & $0.5178$ & $0.3864$ \\
			\end{tabular}
		\end{ruledtabular}
	\end{table}
	\section{Summary and outlook\label{sc:summary}}
	After introducing the spectral representations of the propagators, the BSA, and the BSWF, we converted both the BSE and the definition of the BSWF from identities in the momentum space into those of the Nakanishi spectral functions. Explicitly for scalar particles bound by a scalar-particle exchange interaction, these identities were derived directly from the momentum-space relations without the need of projection operations. We further found that because the coupling constant carried mass dimension, the Naknaishi spectral function of the BSA contained an explicit factor of derivative on the radial spectral variable. The kernel functions we derived for these integration relations were then explicit functions without numerical intermediate steps, which allowed accurate and efficient numerical solutions of the BSE to be obtained. Also because the Nakanishi integral representations captured the discontinuities of Green's functions at boundaries of their holomorphic region, our formulation was suitable for the solution of the bound state structure in Minkowski space. 	
	
	To facilitate the numerical solution of these equations, we proposed an algorithm of adaptive mesh grid for integral equations. This algorithm was subsequently applied together with an iterative solver of integral equations to compute the Nakanishi spectral functions of ground state BSAs in the massive variation of the Wick-Cutkosky model. With a selection of model paramters, our result was in agreement with that in Ref.~\cite{Kusaka:1997xd} with numerical kernel functions. In the future, we plan to solve the BSE with dressed propagators. We also plan to extend our formulation to bound states of fermions through gauge interactions. 
	\begin{acknowledgments}
		S.J. appreciates Prof. Pieter Maris and Prof. James Vary for many discussions on this topic during his assignment at Iowa State University. This work was supported by the U.S. Department of Energy, Office of Science, Office of Nuclear Physics, under Contract No. DE-AC02-06CH11357. We gratefully acknowledge the computing resources provided on Bebop, a high-performance computing cluster operated by the Laboratory Computing Resource Center at Argonne National Laboratory.
	\end{acknowledgments}
	\appendix*
	\section{Adaptive Simpson's rule\label{ss:adaptive_simp}}
	We apply the well-established Simpson's rule to evaluate the integral of a function ${f(x)}$ over a segment ${x\in[a,b]}$ numerically~\cite{abramowitz1965handbook}. At least $5$ quadrature points are required for error estimation. Specifically for the integral on a $5$-point segment, we have 
	\begin{align}
		& \int_{a}^{b}dx\,f(x) \simeq \dfrac{h}{3}\big[ f(a) + 4f(a+h) + 2f(a+2h) \nonumber\\
		& \quad + 4f(a+3h) + f(a + 4h) \big],\label{eq:quad_weights}
	\end{align}
	with ${h=(b-a)/4}$ being the separation of of the grid points. The condition for the error of this segment to be within a desired tolerance of $\epsilon_{\mathrm{tol}}$ is given by
	\begin{align}
		& \bigg\vert \dfrac{h}{3}\big[ -f(a) + 4f(a+h) - 6f(a+2h) + 4f(a+3h) \nonumber\\
		& \quad - f(a + 4h) \big] \bigg\vert < 15 \epsilon_{\mathrm{tol}}.\label{eq:quad_error}
	\end{align}
	No division is needed for the working interval when Eq.~\eqref{eq:quad_error} is satisfied. Otherwise the working interval is divided into $2$ equally spaced intervals, with the same condition applying to both subintervals. To apply this adaptive rule, the initial grid for the integral consists of
	\begin{equation}
		N_{\mathrm{quad}} = 5+4N_{\mathrm{init}} \label{eq:def_N_init}
	\end{equation}
	equally spaced points, with $N_{\mathrm{init}}\in \mathbf{N}$. 
	\bibliography{bse_refs}
\end{document}